\documentclass{article}

\usepackage{arxiv}

\usepackage[utf8]{inputenc} 
\usepackage[T1]{fontenc}    
\usepackage{hyperref}       
\usepackage{url}            
\usepackage{booktabs}       
\usepackage{amsfonts}       
\usepackage{microtype}      
\usepackage{lipsum}
\usepackage{graphicx}
\graphicspath{ {./images/} }

\usepackage{subcaption} 
\usepackage{xspace}
\usepackage{amsmath}
\usepackage{siunitx}
\usepackage{multirow}
\usepackage{adjustbox}

\title{A systematic comparison of hard- and soft-constrained physics-informed molecular machine learning with the Clapeyron equation}

\author{
  Jan Pav\v{s}ek \textsuperscript{1}, \quad Jan G. Rittig \textsuperscript{1}, \quad Alexander Mitsos \textsuperscript{2,1,3*}, \vspace*{2mm}\\
	\textsuperscript{1}{Process Systems Engineering (AVT.SVT), RWTH Aachen University} \\
	\textsuperscript{2}{JARA Center for Simulation and Data Science (CSD)}\\
	\textsuperscript{3}{Institute of Climate and Energy Systems ICE-1: Energy Systems Engineering, Forschungszentrum Jülich GmbH}\\
	\textsuperscript{*}{Corresponding author, \texttt{amitsos@alum.mit.edu}}\\
}

\begin{document}

\maketitle

\begin{abstract}

\noindent In molecular machine learning (ML), significant progress has been made on integrating fundamental thermodynamic relations into ML models. 
Such physics-informed approaches can be categorized as \emph{hard-constrained}, where the thermodynamic relations are embedded in the model architecture, and \emph{soft-constrained}, where relations are included in the training loss. 
While both approaches have been applied to various property prediction tasks, a systematic comparison on experimental data is lacking.
We herein compare hard- and soft-constrained approaches on a numerically challenging case study: the prediction of vapor pressure, saturated liquid and vapor molar volumes, and enthalpy of vaporization as functions of temperature for single-species vapor--liquid equilibrium, related by the exact Clapeyron equation.
Furthermore, soft-constrained approaches in molecular ML typically balance data and physics losses with a fixed weighting factor, requiring careful tuning.
To overcome this, we investigate the augmented Lagrangian method (ALM), well established in constrained optimization.
We find that for the soft-constrained approach, using the ALM improves prediction performance in terms of constraint satisfaction by factor 2 and reduces training epochs compared to using a fixed penalty by 30\%.
Hence, in the soft-constrained approach, the use of the ALM is always recommended.
The hard-constrained approach, which entails additional architectural design choices, can reach on par accuracy with the soft-constrained approach, at even higher thermodynamic consistency, reaching 12 orders of magnitude closer approximation of the Clapeyron equation.
Overall, the hard-constrained approach is most promising when thermodynamic consistency is critical, yet requires tailoring the ML model architecture to the respective thermodynamic relation.
\end{abstract}

\section{Introduction}

\noindent Ensuring adherence to fundamental thermodynamic relations is a key requirement for the practical use of molecular machine learning (ML) models for thermodynamic modeling~\cite{rittig2026molecular,nogueira2026domain,hasse2026artificial,jirasek2023combining}.
Incorporating these relations enhances the thermodynamic consistency, and thus the trustworthiness, of model predictions and, in certain cases, also improves prediction accuracy~\cite{rittig2026molecular,pavvsek2026clapeyron}.
To this end, a variety of approaches have been developed to combine ML and thermodynamics, which we have proposed to categorize into hybrid and physics-informed approaches~\cite{rittig2026molecular}.

In hybrid approaches, ML models are combined with semi-empirical equations such as the Antoine equation or the NRTL model.
The ML models are typically used to predict equation-specific parameters from the molecular structure, targeting species for which such parameters are not readily available.
Within the realm of hybrid models, the coupling of an ML model and a semi-empirical equation can be realized in different ways; see overviews in~\cite{rittig2026molecular, hasse2026artificial}.
Examples for hybrid models are SPT-PC-SAFT~\cite{winter2025understanding}, SPT-NRTL~\cite{winter2023spt_nrtl}, GRAPPA~\cite{hoffmann2025grappa},PUFFIN~\cite{santana2024puffin}, ML-SAFT~\cite{felton2024ml_saft}, Equinet~\cite{alam2026equinet}, TeNNet-SAC~\cite{yang2025physics} and Gibbs-Helmholtz-GNN~\cite{medina2023gibbs}.
Since the semi-empirical equations satisfy certain thermodynamic relations by construction, e.g., NRTL satisfies the Gibbs--Duhem equation, hybrid models inherit this consistency, but also the simplifying assumptions of the underlying equation.

In our categorization~\cite{rittig2026molecular}, physics-informed approaches, on the other hand, introduce thermodynamic knowledge into ML models via fundamental thermodynamic relations, such as the Clapeyron equation, the Gibbs--Duhem equation, or Maxwell relations.
Unlike the semi-empirical models used in hybrid approaches, e.g., UNIFAC or PC-SAFT, these relations do not predict thermodynamic properties by themselves; they only specify how properties must relate to each other to be thermodynamically consistent.
Consequently, they impose less structure on the ML model, but, in turn, introduce no simplifying modeling assumptions~\cite{pavvsek2026deepeosnet}.

Such fundamental thermodynamic relations can be incorporated either as soft- or as hard-constraints.
In soft-constrained thermodynamics-informed ML models, the relations are written into the loss function, so that the parameters of the ML model $\vartheta$ are updated during training to fit both the data and fulfill the fundamental relations as well as possible, hence simultaneously optimizing the accuracy and the consistency of the predictions.
Examples of soft-constrained models include our previous works on Gibbs-Duhem-informed GNNs~\cite{rittig2023gibbs_informed} and Clapeyron-informed GNNs for prediction of single-species vapor liquid equilibria~\cite{pavvsek2026clapeyron}.
For hard-constrained models, sometimes also referred to as thermodynamics-consistent~\cite{rittig2024thermodynamics_consistent}, the thermodynamic relations are directly embedded into the ML model architecture, i.e., the model contains algebraic and/or differential thermodynamic equations.
Examples are our GE-GNN~\cite{rittig2024thermodynamics_consistent}, HANNA~\cite{specht2024hanna, hoffmann2026thermodynamically} and Deep-Helmholtz~\cite{fleck2026deep}. 
The distinction between soft and hard constraints thus refers to the way the thermodynamic knowledge is imposed: soft-constrained models need to learn to satisfy the relations by evaluating the consistency residual during training, whereas hard-constrained models satisfy them for any input by design.

Despite this variety of approaches, few works in molecular ML systematically compare different methods for incorporating thermodynamics into ML models. 
Alam et al.~\cite{alam2026equinet} recently provide a first benchmark between hybrid and hard-constrained physics-informed models for activity coefficient prediction, in which the hybrid models performed better.
However, they acknowledge that their hard-constrained model embeds less information than other hard-constrained models in literature do, e.g., as done by Specht et al.~\cite{specht2024hanna}, which limits the generality of this finding.
Their benchmark also does not include soft-constrained approaches, leaving open whether soft- or hard-constrained approaches are more suitable.
We previously addressed this question for Gibbs--Duhem-consistent activity coefficient prediction~\cite{rittig2024thermodynamics_consistent}, where our initial comparison found the hard-constrained approach to outperform the soft-constrained one.
However, our case study was restricted to simulated data generated with COSMO-based methods and thus free of experimental noise.
Moreover, activity coefficients constitute a comparatively well-posed prediction target, as their values are typically distributed around unity.
Hence, it is unclear how these findings transfer to more complex thermodynamic prediction tasks.
In this work, we therefore expand that analysis by comparing hard- and soft-constrained physics-informed approaches for the prediction of single-species vapor-liquid-equilibrium, using the exact Clapeyron equation to inform the ML models.
We train on experimental data, with the vapor pressure, the vapor and liquid molar volumes, and the enthalpy of vaporization as functions of temperature as prediction targets.
This is a numerically challenging task, as the property values vary over orders of magnitude, thereby reflecting the difficulty of thermodynamic modeling tasks encountered in practice.

Moreover, most works on soft-constrained physics-informed NN training in molecular ML, including our own, combine the data and physics losses ($\mathcal{L}$) via a weighted sum~\cite{rosenberger2022machine,rittig2023gibbs_informed,hammad2025advancements,leenhouts2025thermodynamics,park2025multi,hoffmann2025machine,pavvsek2026clapeyron}:
\begin{equation*}
    \mathcal{L} = \mathcal{L}_\text{data} + \nu \, \mathcal{L}_\text{phys},
\end{equation*}
where the weighting factor $\nu$ is a constant hyperparameter that balances prediction accuracy against thermodynamic consistency.
Note that we use $\nu$ here to denote the fixed penalty factor, instead of $\lambda$ which is commonly used in physics-informed literature, to avoid confusion with notation used in mathematical programming.
Finding a numeric value for $\nu$ that performs this balance reliably throughout training is a major challenge:
Its value strongly affects training performance, may cause non-convergent training, and is highly specific to the prediction task, making hyperparameter optimization challenging.
We herein demonstrate that this problem can be circumvented by using the augmented Lagrangian method (ALM) for robust balancing of physics and data loss.
The ALM, which was introduced in 1969~\cite{hestenes1969multiplier,powell1969method} as an evolution of the quadratic penalty method, is commonly used in gradient-based optimization for constraint handling.
That is, solving constrained optimization problems is converted to solving a series of unconstrained optimization problems instead, accounting for the constraint by means of the Lagrangian.
The ALM has been shown to be well suited for loss-balancing in physics-informed neural network training~\cite{son2023enhanced}, as these training tasks are essentially a series of optimization problems which aim to fulfill a physics equality constraint. 
In fact, a handful of works have indeed already applied the ALM for physics-informed neural network (PINN) training, including Dener et al.~\cite{dener2020training}, Lu et al.~\cite{lu2021physics}, Son et al.~\cite{son2023enhanced}, Hu et al.~\cite{hu2026conditionally} and Deng et al.~\cite{deng2026augmented}, yet the application to molecular ML is missing so far.
In this work, we therefore evaluate the ALM as an alternative to fixed weighting in soft-constrained thermodynamics-informed training.

In summary, this work makes two contributions:
\begin{enumerate}
    \item a systematic comparison of hard- and soft-constrained physics-informed ML for single-species VLE prediction on experimental data;
    \item an evaluation of the ALM as an alternative to fixed weighting in soft-constrained thermodynamics-informed training.
\end{enumerate}
To this end, we compare three approaches: two soft-constrained approaches, based on fixed weighting and on the ALM, respectively, and a hard-constrained approach that embeds the Clapeyron equation in the output head of the model.
From this comparison, we derive practical guidance for physics-informed ML-based thermodynamic modeling.

\section{Methodology}

\noindent We first introduce the soft-constrained and hard-constrained molecular ML model, followed by details on the dataset, and the implementation and hyperparameters.
Notably, we base all models on graph neural networks (GNNs), which are frequently used as end-to-end ML models for molecular property prediction; see overviews in~\cite{reiser2022graph, Rittig_GNNBook.2022}.
As we combine them with the Clapeyron equation, we refer to the models as Clapeyron-GNN in the following.

\subsection{Soft-constrained Clapeyron-GNN}

\begin{figure}[h]  
    \centering
    \includegraphics[trim = 19 25 30 160, clip ,width=\textwidth]{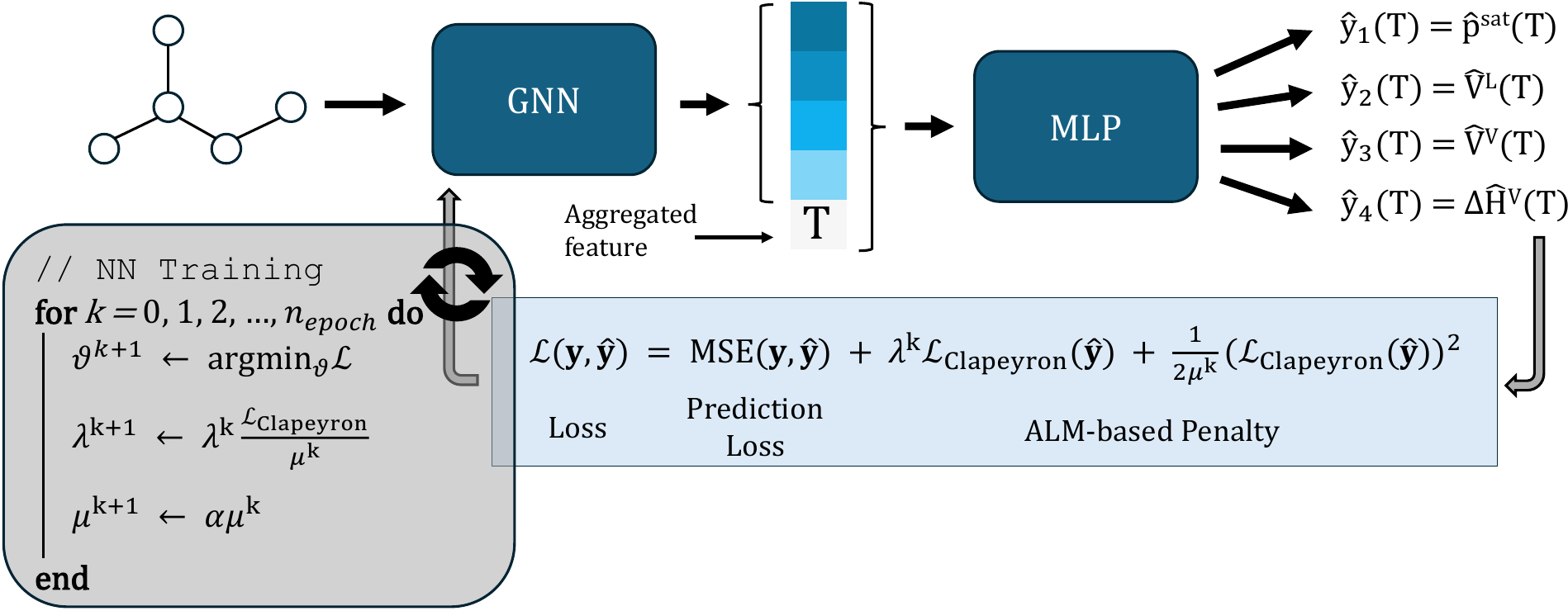}  
    \caption{Schematic illustration ALM-based soft-constrained Clapeyron-informed GNN architecture}
    \label{fig:ci_gnn_arch}
\end{figure}

\noindent The soft-constrained Clapeyron-GNN using the augmented Lagrangian method (ALM) is illustrated in Figure~\ref{fig:ci_gnn_arch}.
We thereby extend our previously developed soft-constrained Clapeyron-GNN using a fixed weighting factor for the Clapeyron loss ($\mathcal{L}_\text{Clapeyron}$) in \cite{pavvsek2026clapeyron} by the ALM, cf. \cite{dener2020training}.

The ALM, introduced in 1969~\cite{hestenes1969multiplier,powell1969method}, is an established method in gradient-based optimization -- also known as mathematical programming--, where it is used for constraint handling. 
Consider equality-only nonlinear continuous optimization problems:
\begin{align*} 
    \min_{x}& \quad f(x)  \\
    s.t.& \quad g_{i}(x) = 0, i \in E 
\end{align*}
where $f(\cdot)$ is the objective function and $g_{i}(\cdot)$ are the (equality) constraint functions. 
The ALM belongs to the family of penalty methods, and even more generally to those that solve the optimization problem as a series of unconstrained problems.
In the earlier quadratic penalty method
\begin{equation*} 
    L(x, \mu) = f(x) + \frac{1}{2\mu}\sum_{i\in E}(g_{i}(x))^{2}
\end{equation*}
the penalty is progressively increased as the parameter is reduced ($\mu \rightarrow 0$).
For each value of the parameter, $L$ is minimized using methods for unconstrained optimization taking the previous point $x$ as the initial guess. 
In the ALM, the equality constraints are similarly included in the objective function:
\begin{equation*} 
    L(x,\lambda, \mu) = f(x) + \sum_{i\in E}\lambda_{i}g_{i}(x) + \frac{1}{2\mu}\sum_{i\in E}(g_{i}(x))^{2}
\end{equation*}
Again, $\mu$ is progressively reduced; the parameter $\lambda$ is updated based on duality theory. Thereby, ALM solves the constrained optimization problem in fewer iterations, and more importantly with more moderate values of $\mu$, thus avoiding ill-conditioned unconstrained minimization problems.
For more details on the ALM and penalty methods, we direct the interested reader to \cite{nocedal2006numerical}.

The majority of (soft-constrained) physics-informed molecular ML publications use the simple yet less effective fixed penalty method~\cite{rosenberger2022machine,hammad2025advancements,leenhouts2025thermodynamics,park2025multi,hoffmann2025machine,rittig2023gibbs_informed,pavvsek2026clapeyron}, where the physics loss is added to the data loss by means of a fixed factor $\nu$ which is chosen through hyperparameter optimization, as opposed to progressively as in ALM and the penalty method of optimization. 
The main drawback of having fix value $\nu$ is that training performance is highly sensitive to the numeric value of $\nu$, which on the one hand makes hyperparameter optimization challenging and on the other can lead to non-convergent training behavior.
To address this, the ALM has been applied in recent years to physics-informed NN training \cite{raissi2019physics}, where the objective function $f(\cdot)$ corresponds to the data loss during training, e.g., mean squared error, and the equality constraints $g_{i}(\cdot)$ correspond to the physics relations, used to inform the ML model \cite{dener2020training,lu2021physics,son2023enhanced,hu2026conditionally,deng2026augmented}.

In our soft-constrained Clapeyron-GNN, the objective function $f(\cdot)$ corresponds to the mean squared error between predicted and true value, and the constraint function $g(\cdot)$ corresponds to the Clapeyron error
\begin{equation*} 
    \mathcal{L}_{\text{Clapeyron}} = \left(\frac{\frac{d\hat{p}^{sat}}{dT} T (\hat{V}^{V} - \hat{V}^{L})}{\hat{\Delta H_{V}}} - 1\right)^{2},
\end{equation*}
which we define in \cite{pavvsek2026clapeyron}.
Hence, the loss function using ALM reads:
\begin{equation*} 
    \mathcal{L}(y,\hat{y}) = MSE(y,\hat{y}) + \lambda\mathcal{L}_{\text{Clapeyron}}(\hat{y}) + \frac{1}{2\mu}(\mathcal{L}_{\text{Clapeyron}}(\hat{y}))^{2},
\end{equation*}
where $\lambda$ and $\mu$ are dynamic hyperparameters, which get updated inside the training loop simultaneously with the neural network weights.
Training with dynamic ALM hyperparameters therefore does not add an outer training loop.
The update steps of the ALM parameters are defined by:
\begin{align*}
    \lambda^{k+1} &\leftarrow \lambda^{k} + \frac{\mathcal{L}_{\text{Clapeyron}}(\hat{y}^{k})}{\mu^{k}} \\
    \mu^{k+1} &\leftarrow \alpha\mu^{k}
\end{align*}
where $k$ refers to the current epoch and $\alpha$, $\lambda^{k=0}$ and $\mu^{k=0}$ are selected through hyperparameter optimization.
Note that these hyperparameters are unique to the ALM-based soft-constrained approach, and are therefore additional to the hyperparameters of the underlying GNN.
In particular, the ALM increases the number of additional hyperparameters from one in the fixed penalty approach, i.e., $\nu$, to three ($\alpha$, $\lambda^{k=0}$ and $\mu^{k=0}$).
However, as the numeric values of the ALM hyperparameters are less specific to the prediction task than the fixed penalty is, 
optimizing them may not always be needed and is less challenging when they are indeed included in hyperparameter optimization.

\subsection{Hard-constrained Clapeyron-GNN}

\begin{figure}[h]  
    \centering
    \includegraphics[trim = 19 95 30 140, clip ,width=\textwidth]{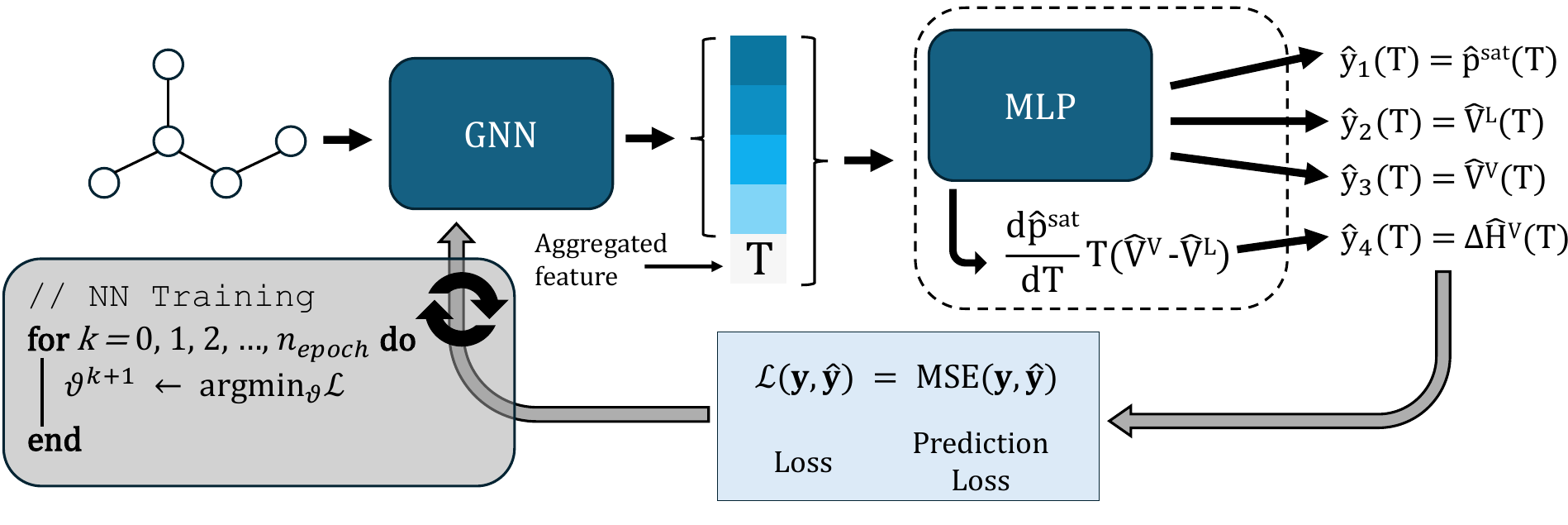}  
    \caption{Schematic illustration hard-constrained Clapeyron-informed GNN architecture}
    \label{fig:cc_gnn_arch}
\end{figure}

\noindent For the hard-constrained Clapeyron-GNN (see Figure~\ref{fig:cc_gnn_arch}), we write the Clapeyron equation directly in the output head, i.e., the multilayer perceptron (MLP), of the model and train purely based on the prediction loss, similar to what we did in our original work on thermodynamics-consistent GNNs \cite{rittig2024thermodynamics_consistent}.
Contrary to that work, where the MLP predicts an intermediate quantity, i.e., the Gibbs excess free energy $G^{E}$, from which measurable output quantities, i.e., the activity coefficients $\gamma$, are obtained by derivation, here we have four measurable prediction targets which are explicitly coupled by the Clapeyron equation.
This requires predicting three of the four properties with the MLP and calculating the fourth explicitly using the Clapeyron equation.
This structural requirement adds additional complexity to the design of the ML architecture, i.e., which property to pick as the calculated one and how to handle its data transformation, which is why we have not used the hard-constrained approach in our previous work~\cite{pavvsek2026clapeyron}.

Here, we select the enthalpy of vaporization to be calculated by the Clapeyron equation, from predictions of the vapor pressure and vapor and liquid molar volume:
\begin{equation*} 
    \Delta \hat{H}_{V} =\frac{d\hat{p}^{sat}}{dT} T (\hat{V}^{V} - \hat{V}^{L})   
\end{equation*}
This selection avoids fractions and integrations inside our hard-constrained model, contributing to numeric stability during training. 
Naturally, the numeric values of the four properties vary across orders of magnitude; thus, all training is performed on log$_{10}$-transformed target values.
This requirement also holds for the calculated $\Delta \hat{H}_{V}$ values, as not transforming them would result in unbalanced loss signals.
However, as the logarithm is only defined on positive values, it needs to be ensured that the values of $\Delta \hat{H}_{V}$, calculated by the Clapeyron equation, are always positive.
As this does not inherently follow from the equation's structure, we perform a ReLU transformation with slope one and intercept zero and add a small positive $\epsilon$ before performing the log$_{10}$-transformation.
Thereby, it is ensured that the values predicted during training for the logarithmic enthalpy are always finite real numbers.
We note that the technically necessary $\epsilon$ addition introduces a small deviation from the Clapeyron equation into the model predictions.
Hence, predictions made by the hard-constraint Clapeyron-GNN can strictly speaking not be referred to as thermodynamics-consistent, which is why we refer to this modeling approach as ``hard-constrained'' following our terminology introduced in~\cite{rittig2026molecular}, rather than ``thermodynamics-consistent'' as used in previous work~\cite{rittig2024thermodynamics_consistent}.
However, given the fact that the $\epsilon$ is small, we expect the deviations to be negligible in practice, such that the predictions made by the hard-constrained Clapeyron-GNN can likely be considered ``near''-consistent, which we analyze in more detail in Section~\ref{sec:res_n_dis}.

\subsection{Dataset \& Training}

\noindent We use the same dataset obtained from the NIST TDE \cite{nist_tde} as in our previous work \cite{pavvsek2026clapeyron}. 
The dataset contains 102,121 datapoints across 879 molecules in a temperature range from 56.75K to 1021K. 
It is worth noting that the dataset is highly unbalanced, as there are significantly more data on the vapor pressure and liquid molar volume, 78,840 and 43,056 respectively, than on the enthalpy of vaporization and vapor molar volume, 1,057 and 2,206 respectively.
A more detailed statistical analysis of the dataset can be found in \cite{pavvsek2026clapeyron}.

For model training, we employ the molecular extrapolation train-test split from our previous work.
That is, we sort the datapoints on 80\% of the molecules in the training set and reserve the data on the remaining 20\% for testing.
For result generation, we perform model training for both models with 10 different seeds. 
Overall metrics are reported as averages with standard deviations across the different seeds.
For specific performance comparison, i.e., in parity plots and individual molecule predictions, the best model out of the ten different seeds for both models is shown.

\subsection{Implementation \& Hyperparameters}

\noindent The hard- and soft-constrained models are implemented in our open-source \href{https://gitlab.git.nrw/rwth-avt-svt/public/gmolprop}{GMoLprop} framework, which is based on PyTorch and PyTorch Geometric \cite{FeyLenssen2019pytorchgeom}.
For all three models, hyperparameter optimization is performed separately using grid search, where hyperparameter optimization for the soft-constrained model with fixed penalty has been performed in our previous work~\cite{pavvsek2026clapeyron}. 
For the hard-constrained Clapeyron-GNN, the optimized hyperparameters are batch size $\in$ \{64, 128\}, fingerprint dimension $\in$ \{64, 128\}, activation function $\in$ \{LeakyReLU, SiLU\} and transformation type of enthalpy prediction $\in$ \{norm, log$_{10}$\}.
Grid search results in a batch size of 64, a fingerprint dimension of 128, the SiLU activation function and the log$_{10}$ transformation for the enthalpy predictions.
For the soft-constrained Clapeyron-GNN, grid search yields the same values for the batch size, the fingerprint dimension and the activation function.
Beyond that, for the ALM-based soft-constrained Clapeyron-GNN we additionally optimize $\mu^{k=0} \in$ \{10, 1, 0.1\}, $\lambda^{k=0} \in$ \{0.001, 0.01, 0.1\} and $\alpha \in$ \{0.72, 0.92\}, resulting in a value for $\mu^{k=0}$ of 10, $\lambda^{k=0}$ of 0.001 and $\alpha$ of 0.92.
The optimized hyperparameters for the fixed-penalty soft-constrained Clapeyron-GNN are taken from~\cite{pavvsek2026clapeyron}, with a batch size of 64, a fingerprint dimension of 128, LeakyReLU as the activation function and a weighting factor of 0.1 for the Clapeyron-GNN.

\section{Results and Discussion} \label{sec:res_n_dis}

\noindent We first compare the overall performance of the soft-constrained Clapeyron-GNN with fixed penalty and ALM and the hard-constrained one, followed by an analysis of individual molecule predictions.

\subsection{Model Comparison}

\begin{table}[t!]
\centering
\caption{Performance metrics, root mean squared error (RMSE), mean absolute error (MAE), and coefficient of determination ($\text{R}^{2}$) evaluated on the logarithmic scale on the test set for the two different models and for all four properties, and Clapeyron error evaluated on the test set. Large font values are averages and small fonts standard deviations. Results for the soft-constrained model with fixed penalty taken from our previous work~\cite{pavvsek2026clapeyron}}
\label{tab:perform_met}
\begin{adjustbox}{max width=1.0\textwidth,center}
\begin{tabular}{c|c c c|c c c|c c c}

\multirow{2}{*}{Property} & \multicolumn{3}{c|}{Soft-constrained with fixed penalty~\cite{pavvsek2026clapeyron}} & \multicolumn{3}{c|}{Soft-constrained with ALM} & \multicolumn{3}{c}{Hard-constrained}  \\
& RMSE & MAE & $\text{R}^{2}$ & RMSE & MAE & $\text{R}^{2}$ & RMSE & MAE & $\text{R}^{2}$ \\
 \hline

$p^{sat}\text{(T)}$ &\multirow{2}{*}{0.26 {\tiny $\pm$ 0.019}} & \multirow{2}{*}{0.14 {\tiny $\pm$ 0.0065}} & \multirow{2}{*}{0.97 {\tiny $\pm$ 0.0040}}& \multirow{2}{*}{0.27 {\tiny $\pm$ 0.020}} & \multirow{2}{*}{0.13 {\tiny $\pm$ 0.0081}} & \multirow{2}{*}{0.97 {\tiny $\pm$ 0.0042}} & \multirow{2}{*}{0.29 {\tiny $\pm$ 0.021}} & \multirow{2}{*}{0.14 {\tiny $\pm$ 0.0079}} & \multirow{2}{*}{0.97 {\tiny $\pm$ 0.0045}} \\
{\small(\#78,840)} & & & & & & \\
\hline
$V^{V}\text{(T)}$&\multirow{2}{*}{0.18 {\tiny $\pm$ 0.019}} & \multirow{2}{*}{0.14 {\tiny $\pm$ 0.016}} & \multirow{2}{*}{0.95 {\tiny $\pm$ 0.011}}& \multirow{2}{*}{0.19 {\tiny $\pm$ 0.039}} & \multirow{2}{*}{0.15 {\tiny $\pm$ 0.023}} & \multirow{2}{*}{0.94 {\tiny $\pm$ 0.030}} & \multirow{2}{*}{0.19 {\tiny $\pm$ 0.036}} & \multirow{2}{*}{0.14 {\tiny $\pm$ 0.023}} & \multirow{2}{*}{0.94 {\tiny $\pm$ 0.025}} \\
{\small(\#2,206)} & & & & & & \\
\hline
$V^{L}\text{(T)}$ &\multirow{2}{*}{0.048 {\tiny $\pm$ 0.0050}} & \multirow{2}{*}{0.029 {\tiny $\pm$ 0.0046}} & \multirow{2}{*}{0.95 {\tiny $\pm$ 0.012}}& \multirow{2}{*}{0.049 {\tiny $\pm$ 0.0074}} & \multirow{2}{*}{0.027 {\tiny $\pm$ 0.0020}} & \multirow{2}{*}{0.94 {\tiny $\pm$ 0.019}} & \multirow{2}{*}{0.069 {\tiny $\pm$ 0.029}} & \multirow{2}{*}{0.030 {\tiny $\pm$ 0.0046}} & \multirow{2}{*}{0.87 {\tiny $\pm$ 0.12}} \\
{\small(\#43,056)} & & & & & & \\
\hline
$\Delta H_{V}\text{(T)}$&\multirow{2}{*}{0.10 {\tiny $\pm$ 0.023}} & \multirow{2}{*}{0.075 {\tiny $\pm$ 0.018}} & \multirow{2}{*}{0.85 {\tiny $\pm$ 0.063}}& \multirow{2}{*}{0.12 {\tiny $\pm$ 0.026}} & \multirow{2}{*}{0.086 {\tiny $\pm$ 0.024}} & \multirow{2}{*}{0.81 {\tiny $\pm$ 0.078}} & \multirow{2}{*}{0.40 {\tiny $\pm$ 0.58}} & \multirow{2}{*}{0.13 {\tiny $\pm$ 0.15}} & \multirow{2}{*}{<0} \\
{\small(\#1,057)} & & & & & & \\
\noalign{\hrule height 1.5pt}
$\mathcal{L}_{\text{Clapeyron}}$& \multicolumn{3}{c|}{$6.9 \times 10^{-3}$ {\tiny $\pm$ $5.1 \times 10^{-3}$}}& \multicolumn{3}{c|}{$3.5 \times 10^{-3}$ {\tiny $\pm$ $2.6\times 10^{-3}$}} & \multicolumn{3}{c}{\textbf{$9.6 \times 10^{-15}$}
 {\tiny $\pm$ $2.3 \times 10^{-15}$}}  \\
\hline

\end{tabular}
\end{adjustbox}
\end{table}

\noindent In Table~\ref{tab:perform_met}, we report the performance metrics across the four target properties, vapor pressure, vapor and liquid molar volume and enthalpy of vaporization for the hard-constrained Clapeyron-GNN and the soft-constrained one with ALM and fixed penalty, respectively.
The results for the latter are taken from our previous study, where we use the same dataset~\cite{pavvsek2026clapeyron}.
The root mean square error (RMSE), mean absolute error (MAE), and coefficient of determination (R$^{2}$) are given for the prediction accuracy, i.e., the fit between experimental and predicted property value, whereas $\mathcal{L}_{\text{Clapeyron}}$ corresponds to the deviation from the Clapeyron equation. 

\subsubsection{Fixed penalty vs ALM soft-constrained Clapeyron-GNN}

We first compare the two soft-constrained approaches, namely the fixed penalty and the ALM.
Overall, the prediction accuracy is on par between both models.
In terms of the Clapeyron error, however, training with the ALM yields an average improvement of a factor 2 compared to the fixed penalty approach, i.e., from $6.9\times 10^{-3}$ with the fixed penalty to $3.5\times 10^{-3}$ with the ALM.
This reflects the capabilities the ALM shows in gradient based optimization for achieving better constraint satisfaction.
Beyond the closer approximation of the Clapeyron equation, the ALM also reduces the number of epochs required for training.
To quantify this, we define the number of epochs after which training has converged as the first epoch where the relative decrease of the overall loss from one epoch to the next is less than 0.2\%.
Over the ten seeds, the average number of epochs until convergence by this metric is 27 when training with ALM compared to 40 when using the fixed penalty term.
Hence, in this setting, the ALM requires 30\% fewer epochs than the fixed penalty approach for training to converge.
Apart from the better constraint approximation and faster convergence, training with the ALM also overcomes the high sensitivity of training convergence towards the physics penalty factor observed in soft-constrained models with a fixed factor.
This is in part also due to the easier hyperparameter optimization for the ALM, as its hyperparameter values are less task-specific and training performance is less sensitive to its numeric values.
Using the ALM, soft-constrained thermodynamics-informed ML models therefore become an alternative with improved Clapeyron approximation and more robust training behavior.

\subsubsection{Soft-constrained vs. hard-constrained Clapeyron-GNN}

Comparing the hard-constrained Clapeyron-GNN to the soft-constrained one with ALM, the prediction performance is on par between both models, in particular for the vapor pressure and the vapor molar volume, where the differences are not significant.
For the liquid molar volume and the enthalpy of vaporization, however, the soft-constrained approach on average outperforms the hard-constrained one. 
While the difference is comparatively small for the liquid molar volume, with both models showing similar standard deviations in the performance metrics, for the enthalpy of vaporization, the difference is more pronounced.
In fact, for this target property, seven out of the ten models achieve performance on par with the soft-constrained one, but the remaining three deviate largely, resulting in an average R$^{2}$ smaller than zero.
We attribute this largely to the structure of the model, where the enthalpy predictions are explicitly calculated by the Clapeyron equation.
The ReLU mapping in particular, which is technically necessary for the log$_{10}$ transformation of the enthalpy predictions for a balanced data loss signal, adds difficulty during training, since, for the entire area of the NN-parameter space where the Clapeyron equation yields a negative result for the enthalpy, the gradient information for the enthalpy prediction is zero.
This may make it difficult for gradient-based training to reach regions of the parameter space that yield physically meaningful enthalpies of vaporization, rendering training sensitive to the weight initialization, i.e., to the random seed.
We note that there are other formulations of a hard-constrained Clapeyron-GNN, which may not rely on a ReLU transformation, and may therefore not exhibit this problem.
However, in our trials, these came with their own challenges in terms of numeric stability, such that we believe our approach here for the hard-constrained Clapeyron-GNN to be the most promising one.

A further factor that might affect the enthalpy prediction in the hard-constrained model may be the distribution of loss signals between the four prediction targets.
That is, if only data for the other properties are available, the prediction loss in terms of the MSE does not contain a loss term for the enthalpy.
In the hard-constraint approach, this results in most datapoints not containing a loss signal for the enthalpy at all, which makes training more challenging.
Although the distribution of data losses is the same in the soft-constrained approach, the Clapeyron loss, which is obtained purely from the predictions and therefore provides a loss signal at every datapoint for all four properties, may have a balancing effect.
However, it is unclear if this also helps with prediction accuracy or only affects consistency of predictions.
It is also worth noting that, in general, data scarcity does not seem to pose a significant challenge to either of the two models.
Both achieve good prediction performance for the vapor molar volume, for which the number of available datapoints lies in the same order of magnitude as for the enthalpy of vaporization.

With the given approach, the soft-constrained Clapeyron-GNN is the more robust model in terms of weight initialization, with predictive performance on par with or above the hard-constrained one.
In terms of thermodynamic consistency, the picture is different.
The hard-constrained Clapeyron-GNN outperforms the soft-constrained one in terms of approximation of the Clapeyron equation on the test set by twelve orders of magnitude (see Table~\ref{tab:perform_met}).
As the hard-constrained Clapeyron-GNN has the Clapeyron equation built into its output head, all predictions adhere to the constraint.
This is contrary to the soft-constrained approach, where the weights must be adapted during training to fit the equation as closely as possible.
Hence, in the soft-constrained approach, good generalization needs to be achieved not only in terms of prediction performance, but also in terms of adherence to the Clapeyron equation.
In the hard-constrained model, adherence to the Clapeyron equation is independent of generalization capabilities.
As we noted above, however, in the hard-constrained case we rely on a small $\epsilon$ addition for numeric stability, which also introduces a small inconsistency.
This is the reason why, also for the hard-constrained model the Clapeyron error $\mathcal{L}_{\text{Clapeyron}}$ does not reach zero.
However, a deviation in the order of magnitude of $10^{-15}$ is negligible.
Notably, also an average Clapeyron error of $3.5\times 10^{-3}$ as observed for the soft-constrained Clapeyron-GNN is expected to be acceptable for most thermodynamic modeling applications.
In practice, the exact tolerance ultimately depends on the specific modeling task.

Finally, our findings here also align with our previous work for activity coefficient prediction~\cite{rittig2024thermodynamics_consistent}.
In that work, hard- and soft-constrained models are on par in terms of predictive performance. 
In terms of thermodynamic consistency, the hard-constrained model outperforms the soft-constrained one, exactly adhering to the embedded thermodynamic relation.
Here, we find the same behavior, but additionally we observe the sensitivity to weight initialization in the hard-constrained approach, which we did not observe in our previous work~\cite{rittig2024thermodynamics_consistent}.
A possible explanation for this is that in their setting, the prediction targets are well distributed and therefore require neither a log$_{10}$ transformation nor a ReLU layer after passing the thermodynamic equation.
In our case, this need is inherent to the more challenging case study, where numeric values are spread out over orders of magnitude.
As this is representative of most thermodynamic modeling tasks, we expect hard-constrained NN training for thermodynamic prediction tasks in practice to be in general more challenging than soft-constrained NN training.

\begin{figure}[h]
    \centering

    \begin{subfigure}[t]{0.47\textwidth}
        \centering
        \includegraphics[width=\linewidth]{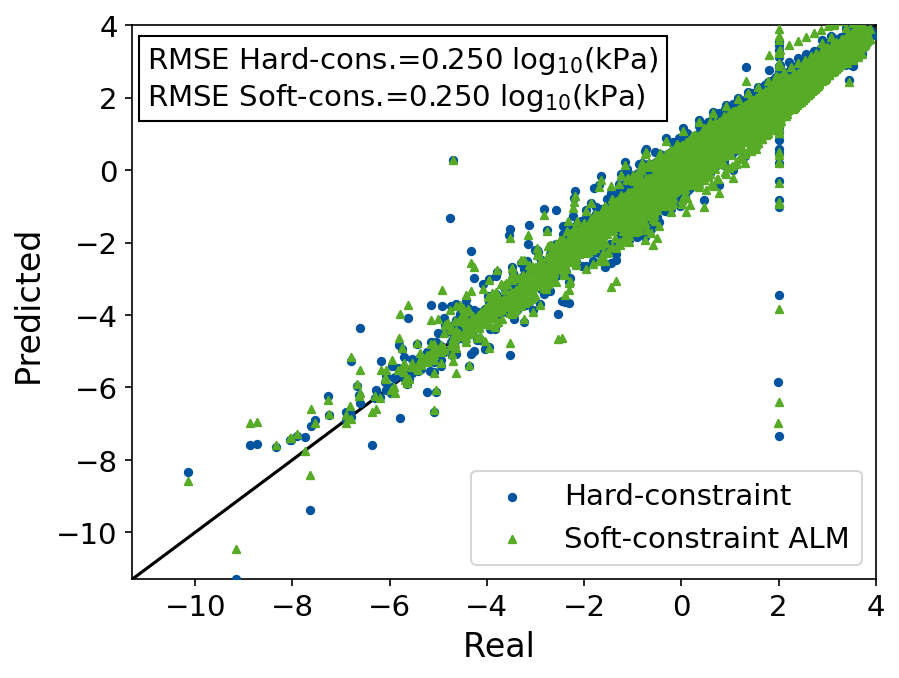}
        \caption{Vapor pressure}
    \end{subfigure}
    \hfill
    \begin{subfigure}[t]{0.47\textwidth}
        \centering
        \includegraphics[width=\linewidth]{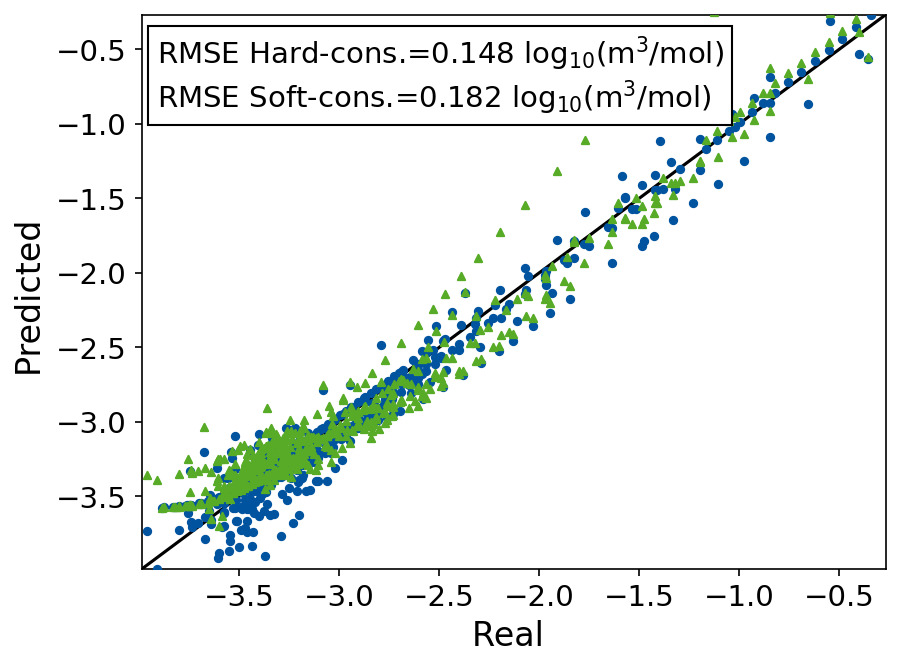}
        \caption{Vapor molar volume}
    \end{subfigure}

    \par\medskip
    
    \begin{subfigure}[t]{0.47\textwidth}
        \centering
        \includegraphics[width=\linewidth]{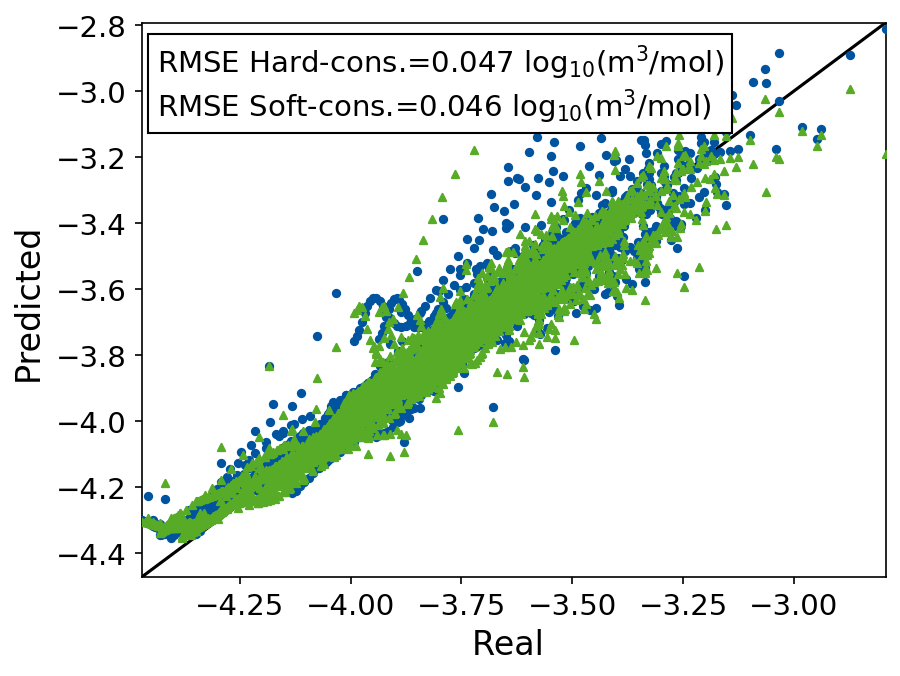}
        \caption{Liquid molar volume}
    \end{subfigure}
    \hfill
    \begin{subfigure}[t]{0.47\textwidth}
        \centering
        \includegraphics[width=\linewidth]{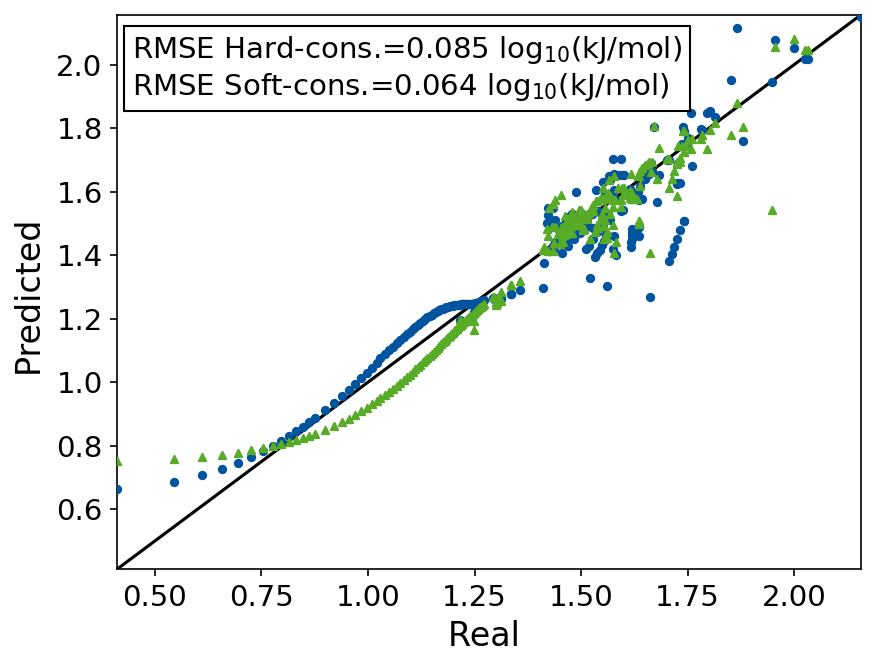}
        \caption{Enthalpy of vaporization}
    \end{subfigure}

    \par\medskip

    \caption{Parity plots of test set hard-constrained Clapeyron-GNN and soft-constrained with ALM}
    \label{fig:parity}
\end{figure}

To analyze the differences in prediction accuracy between the hard-constrained Clapeyron-GNN and the soft-constrained model with the ALM, in Figure~\ref{fig:parity}, we show the parity plots of the test set with the best model of the ten seeds for the four different properties for both models.
The soft-constrained model with fixed penalty, which is on par in prediction accuracy for all four properties with the soft-constrained model with ALM, is not shown for clarity.
For the vapor pressure and the liquid molar volume (see Figures \ref{fig:parity} (a) and (c)), respectively), the performance is on par, albeit with more scatter around the diagonal in the liquid molar volume for both models.
In the vapor molar volume (see Figure \ref{fig:parity} (b)), the overall level of scatter is similar in both models, with the soft-constrained one showing individual trends which deviate significantly from the diagonal, in particular in the region of high numeric values, where the model overestimates the true value.
In the region of low numeric values, the soft-constrained model overestimates the real values, whereas the hard-constrained one underestimates. 
This underlines that for both models the borders of the domain bear the largest challenge for accurate prediction.
For the enthalpy of vaporization the hard-constrained approach exhibits higher scatter than the soft-constrained one.
This is likely due to the fact that in the hard-constrained GNN the enthalpy predictions follow directly from the predictions of the other four properties.
Hence, overfitting to noise in the other three properties directly impacts the enthalpy predictions.
In the soft-constrained model, this is not as pronounced, as adding the thermodynamic relation in the loss term means that deviations from it affect all output predictions equally, such that overfitting for one property is less likely to impact another prediction target.
For hard-constrained models, this spillover effect of overfitting can only happen in a structure where the thermodynamic equation relates directly measurable quantities and one is calculated explicitly from the others.
In the case where intermediate quantities are predicted, and measurable quantities are derived from them through differential equations, as in the GE-GNN \cite{rittig2024thermodynamics_consistent} or in HANNA \cite{specht2024hanna}, the risk of overfitting one property target impacting the prediction performance of another is less pronounced. 

\subsection{Prediction for individual molecules}

\begin{figure}[p]
    \vspace*{-2cm}
    \centering

    \begin{subfigure}[t]{0.4\textwidth}
        \centering
        \includegraphics[width=\linewidth]{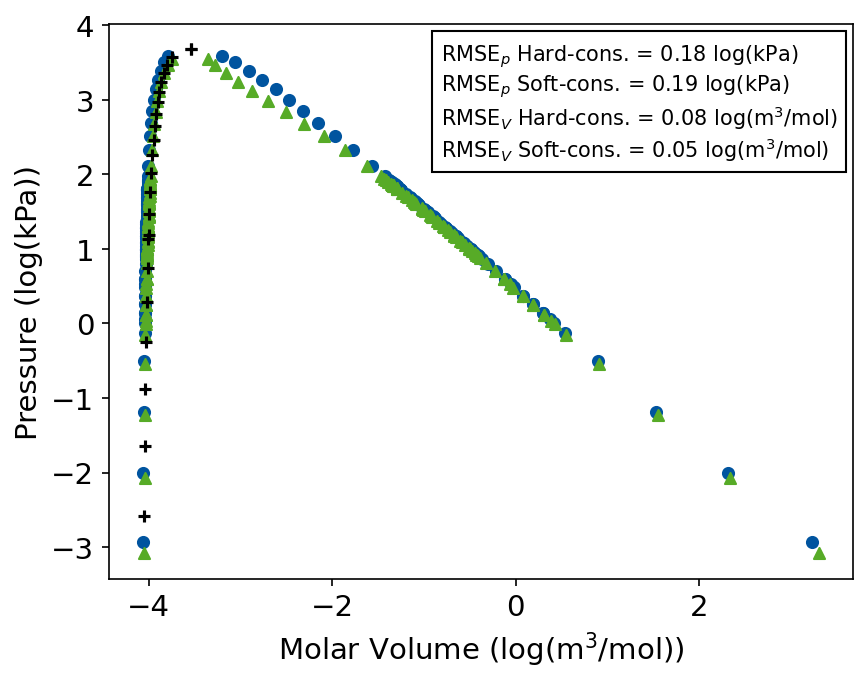}
        \caption{3-Methylthiophene}
    \end{subfigure}
    \hfill
    \begin{subfigure}[t]{0.4\textwidth}
        \centering
        \includegraphics[width=\linewidth]{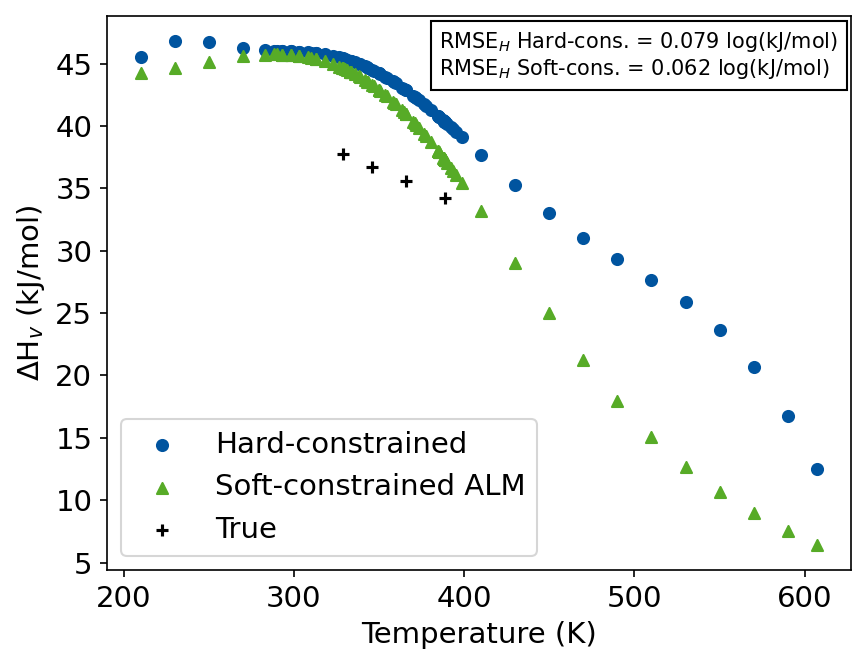}
        \caption{3-Methylthiophene}
    \end{subfigure}

    \par\medskip

    \begin{subfigure}[t]{0.4\textwidth}
        \centering
        \includegraphics[width=\linewidth]{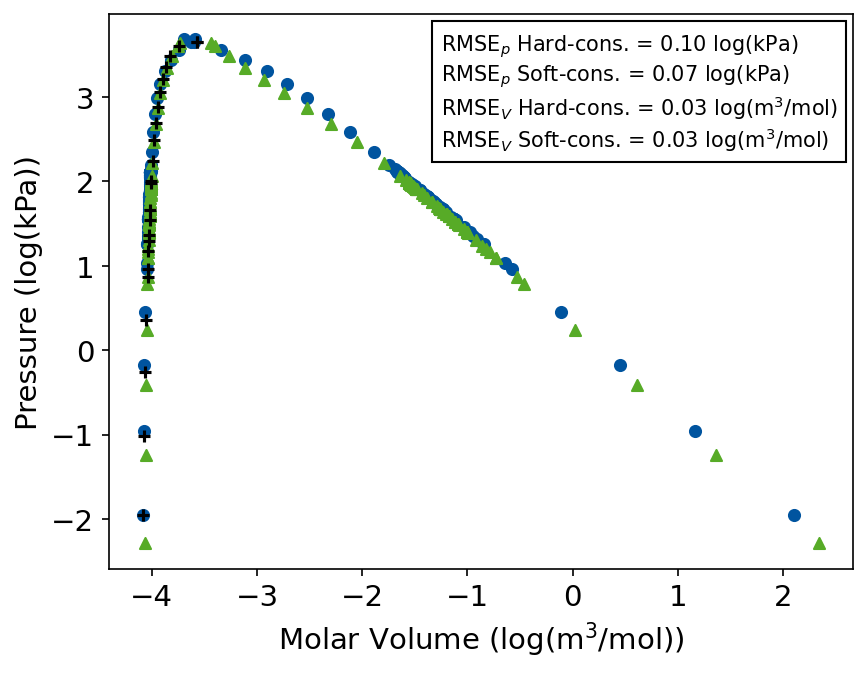}
        \caption{2-Bromopropane}
    \end{subfigure}
    \hfill
    \begin{subfigure}[t]{0.4\textwidth}
        \centering
        \includegraphics[width=\linewidth]{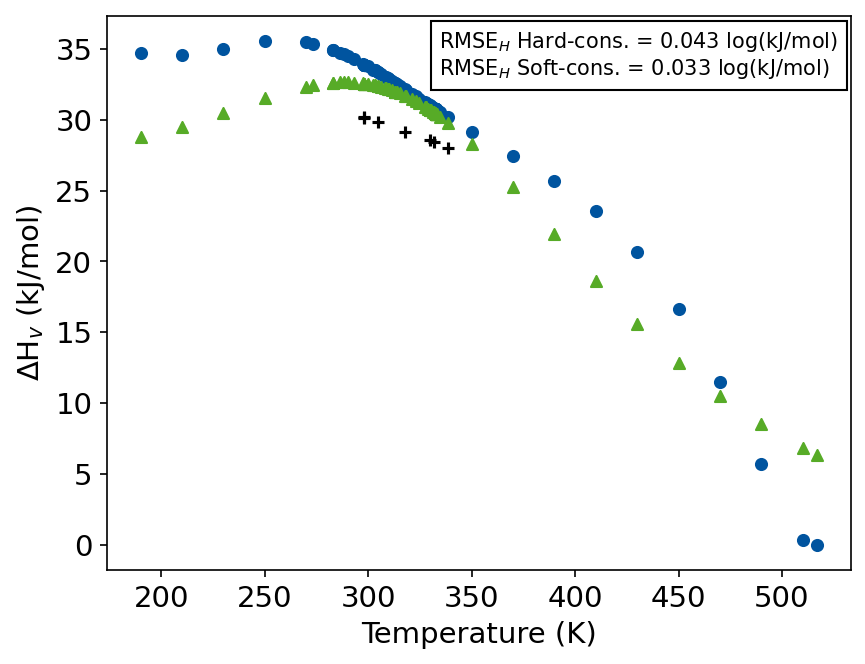}
        \caption{2-Bromopropane}
    \end{subfigure}

    \par\medskip

    \begin{subfigure}[t]{0.4\textwidth}
        \centering
        \includegraphics[width=\linewidth]{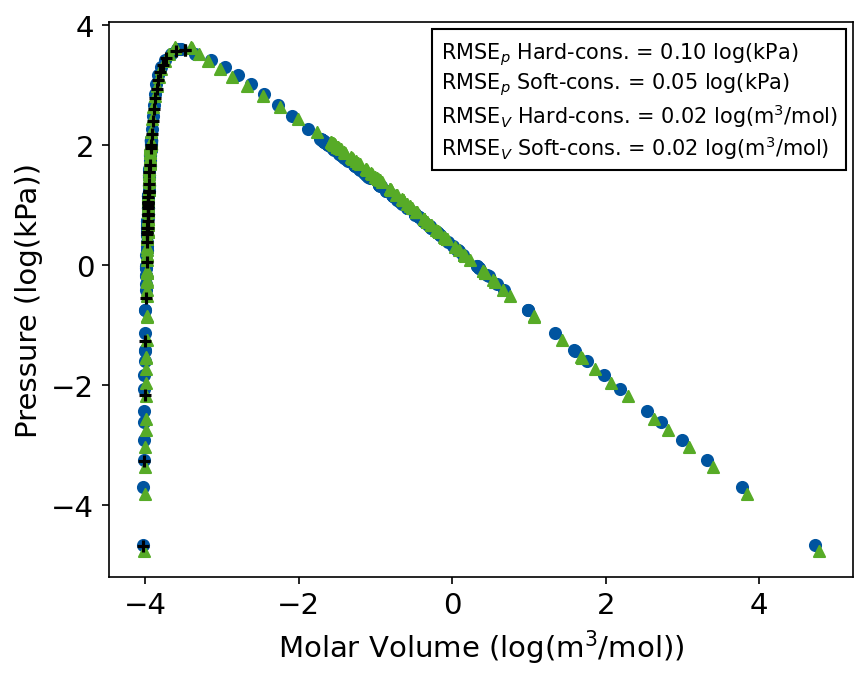}
        \caption{1-Bromobutane}
    \end{subfigure}
    \hfill
    \begin{subfigure}[t]{0.4\textwidth}
        \centering
        \includegraphics[width=\linewidth]{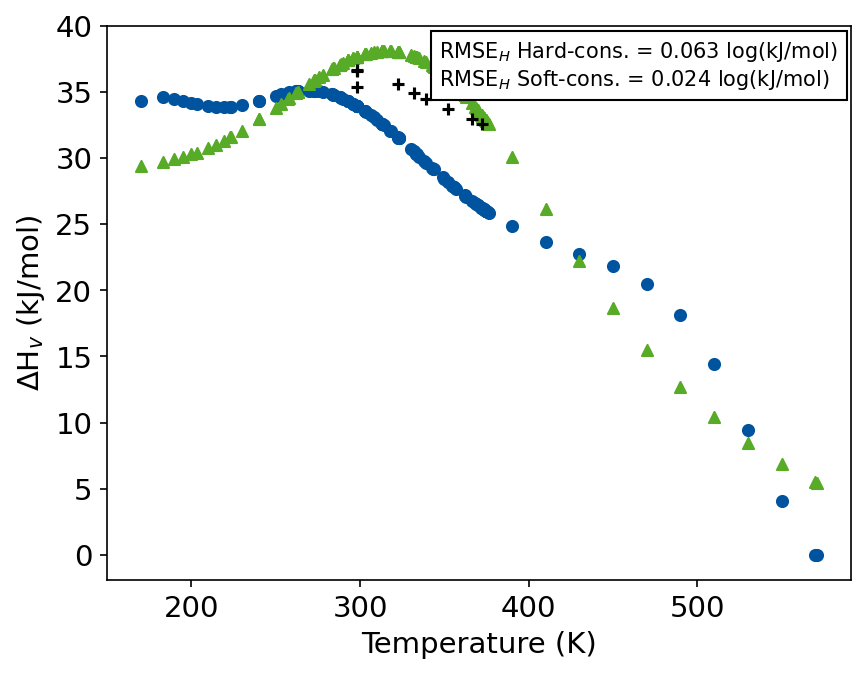}
        \caption{1-Bromobutane}
    \end{subfigure}

    \par\medskip

    \begin{subfigure}[t]{0.4\textwidth}
        \centering
        \includegraphics[width=\linewidth]{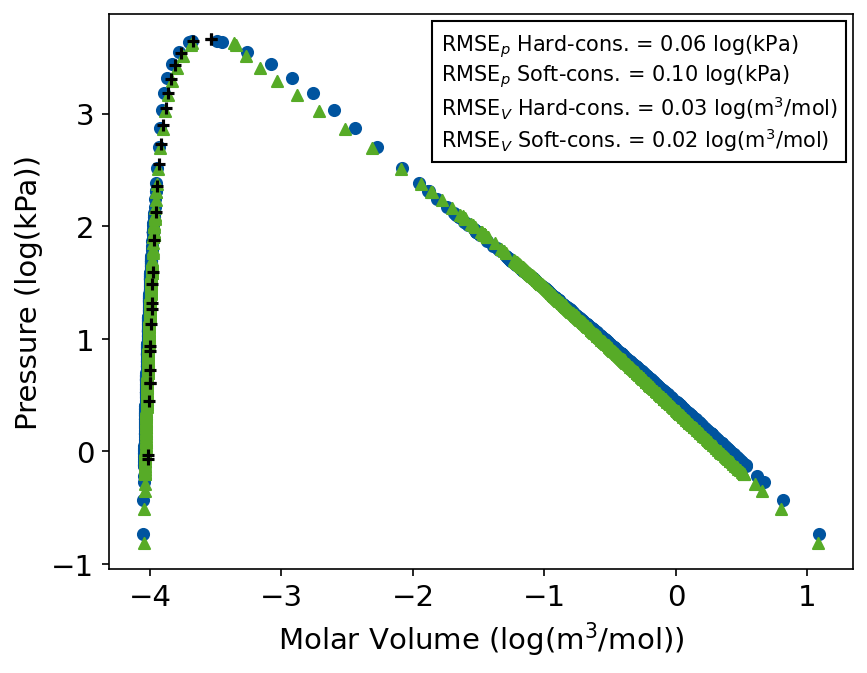}
        \caption{Piperidine}
    \end{subfigure}
    \hfill
    \begin{subfigure}[t]{0.4\textwidth}
        \centering
        \includegraphics[width=\linewidth]{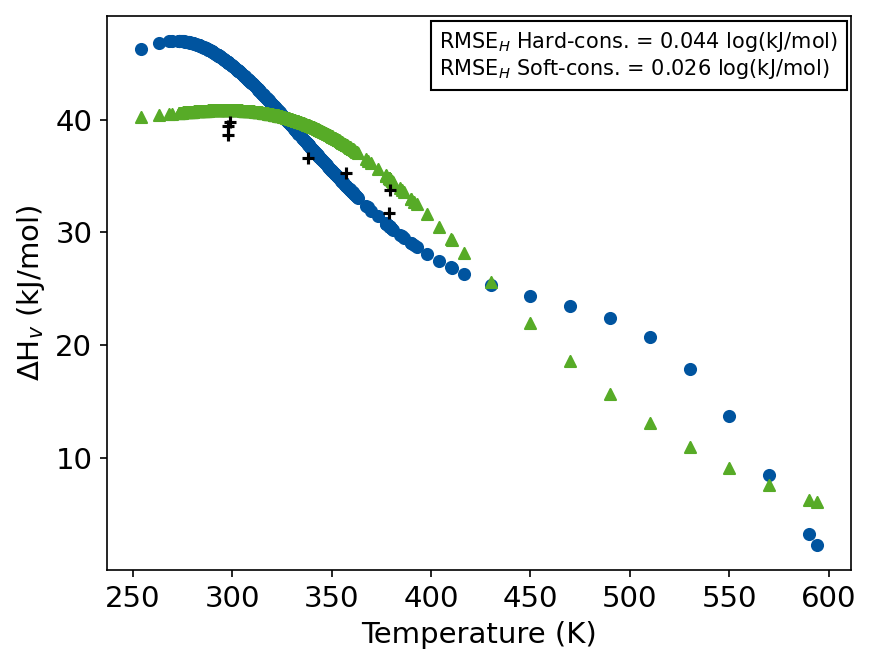}
        \caption{Piperidine}
    \end{subfigure}

    \caption{p(V)-plots and $\Delta$H$_{\text{v}}$(T)-plots for four exemplary molecules of the test set: experimental data in black crosses, hard-constrained in blue dots, and soft-constrained Clapeyron-GNN with ALM in green dots.}
    \label{fig:ind_mol_example}
\end{figure}

To further compare the performance of the hard-constrained Clapeyron-GNN with the soft-constrained one with ALM, the p(V)-plots and $\Delta$H$_{\text{v}}$(T)-plots of four molecules are shown here (see Figure \ref{fig:ind_mol_example}).
The selection of molecules corresponds to the ones shown in our previous work~\cite{pavvsek2026clapeyron}, enabling direct comparison.
In the p(V)-plots, it can be observed that the hard-constrained model tends to predict higher numeric values for the gas phase close to the critical point.
This is particularly pronounced for 3-Methylthiophene and Piperidine. 
As the RMSE of the hard-constrained model is lower for the vapor pressure but higher for the molar volumes than the RMSE of the soft-constrained one for these two molecules, it is not clear which model follows the true trend more closely in this region.
Overall, both models capture the trend in the data well, including the region close to the critical point, which is inherently difficult. 

In the $\Delta$H$_{\text{v}}$(T) plots, the soft-constrained model overall fits the data better, which is reflected in lower RMSE values.
However, the hard-constrained model follows the expected trend better, which is a monotonically decreasing function in T that becomes zero at the critical temperature.
Additionally, the hard-constrained model does not fulfill this trend perfectly, with some non-monotonic regions in 1-Bromobutane and Piperidine.
However, it does follow it more closely, in particular for 2-Bromopropane and 1-Bromobutane, where the soft-constrained model exhibits a very pronounced local maximum, which is not physical.

\begin{figure}[h!]
    \centering

    \begin{subfigure}[t]{0.47\textwidth}
        \centering
        \includegraphics[width=\linewidth]{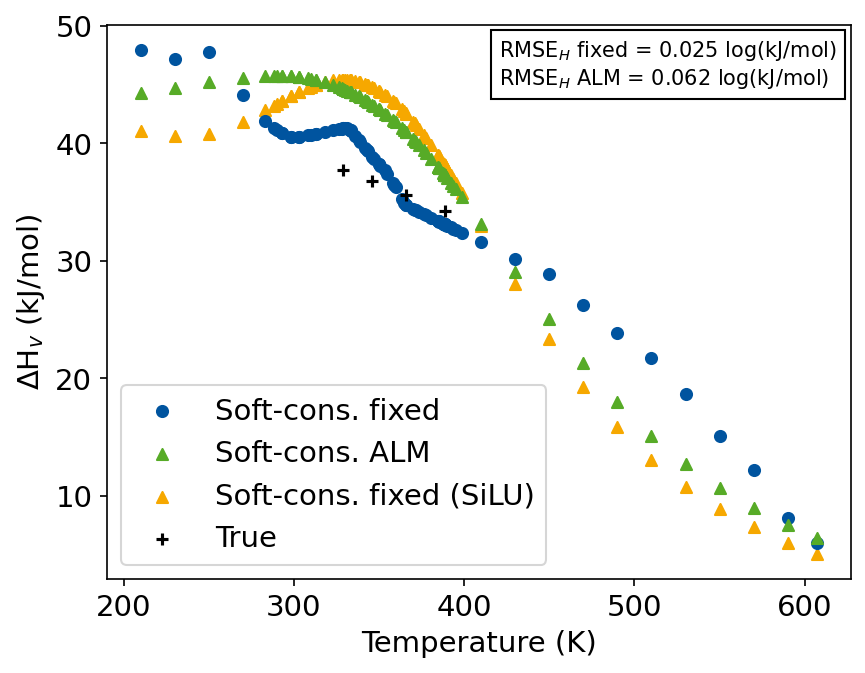}
        \caption{3-Methylthiophene}
    \end{subfigure}
    \hfill
    \begin{subfigure}[t]{0.47\textwidth}
        \centering
        \includegraphics[width=\linewidth]{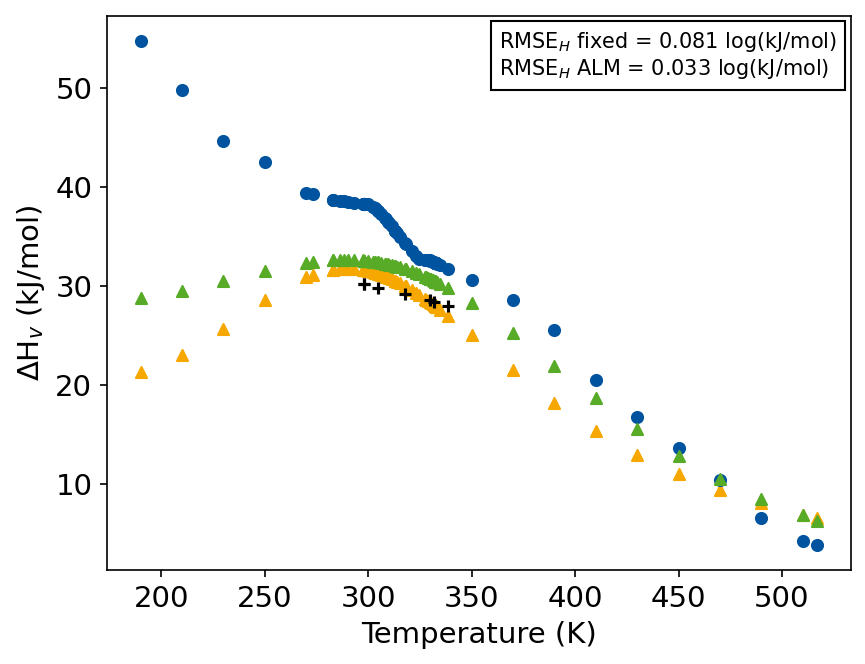}
        \caption{2-Bromopropane}
    \end{subfigure}

    \par\medskip
    
    \begin{subfigure}[t]{0.47\textwidth}
        \centering
        \includegraphics[width=\linewidth]{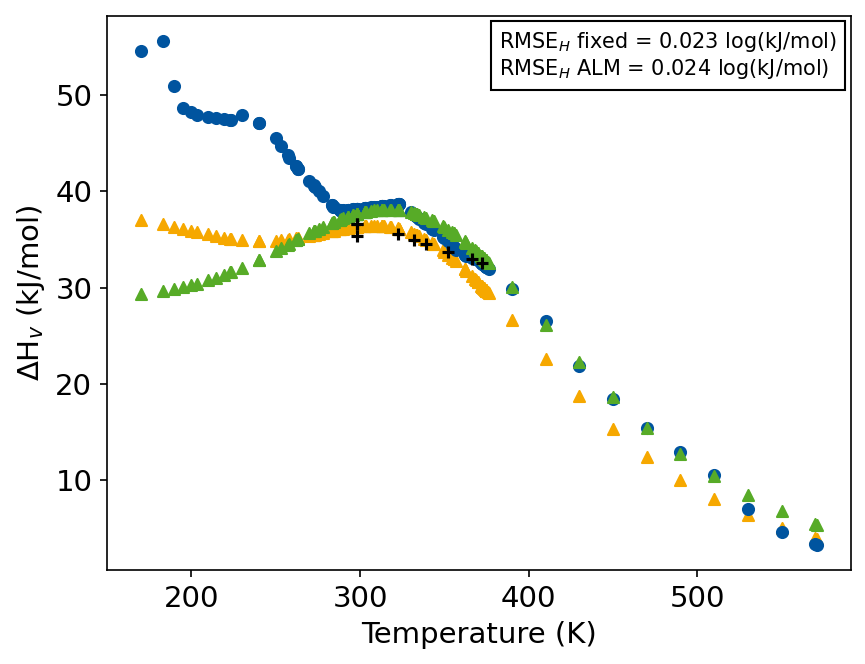}
        \caption{1-Bromobutane}
    \end{subfigure}
    \hfill
    \begin{subfigure}[t]{0.47\textwidth}
        \centering
        \includegraphics[width=\linewidth]{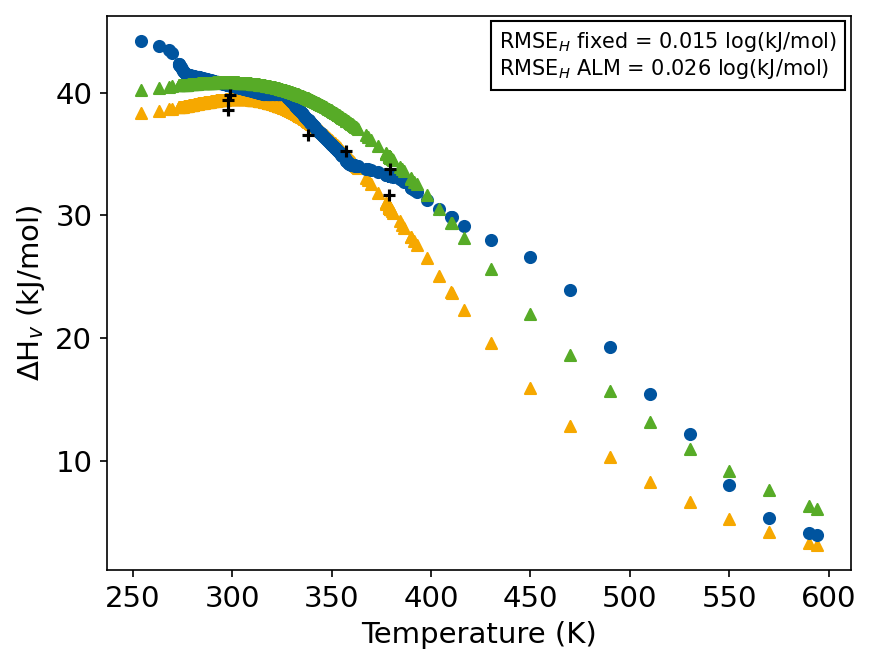}
        \caption{Piperidine}
    \end{subfigure}

    \par\medskip

    \caption{$\Delta$H$_{\text{v}}$(T)-plots for four exemplary molecules of the test set comparing the soft-constrained approach with ALM (green) with the fixed penalty (blue), and fixed penalty using SiLU (yellow). Data for the fixed penalty taken from~\cite{pavvsek2026clapeyron}}
    \label{fig:h_fixed_ALM}
\end{figure}

To qualitatively analyze the difference in adherence to the Clapeyron equation between the soft-constrained approach with ALM and with a fixed penalty, we show the $\Delta$H$_{\text{v}}$(T) plots for individual molecules in Figure~\ref{fig:h_fixed_ALM}.
Note that for the soft-constrained approach with fixed penalty, we show two variants, the blue dots are the predictions of the model using LeakyReLU as activation function, which is the optimized value of the hyperparameter, and the yellow triangles use SiLU as activation function instead.
It is clearly visible that the soft-constrained model trained with the ALM follows the expected functional form of the enthalpy, i.e. a monotonically decreasing function in T, as described above, more closely. 
For the soft-constrained model with a fixed penalty and LeakyReLU, i.e., the model with optimal hyperparameters, yields non-smooth predictions in the enthalpy, which is visible in all four molecules shown, which can be explained by the non-smooth form of the activation function.
When using SiLU, i.e., a smooth activation function, for the soft-constrained approach with a fixed factor for the physics loss during training, we find pronounced local maxima in all four example molecules.
For the model trained with the ALM, this is only the case for 1-Bromobutane and only slightly for 2-Bromopropane.
Hence, training with the ALM provides a significantly better approximation of the thermodynamically expected functional, while reaching on par data fitting.
Therefore, the benefit of ALM-based soft-constrained training over training with a fixed physics penalty factor in terms of thermodynamic consistency can not only be seen quantitatively in the Clapeyron errors (see Table \ref{tab:perform_met}) but also qualitatively in the $\Delta$H$_{\text{v}}$(T) plots.

\section{Conclusion}
\label{sec:conc}

\noindent We benchmark hard- and soft-constrained physics-informed molecular ML on a representative thermodynamic case study for the prediction of single-species vapor liquid equilibrium.
We find both approaches to yield comparable prediction performance, with closer adherence to the Clapeyron equation achieved by the hard-constrained model.
Hence, the hard-constrained approach is most suitable for thermodynamic modeling tasks in practice where thermodynamic consistency is a key requirement. 
Applying a hard-constraint model to a new thermodynamic modeling task, however, requires adapting its architecture to the specific thermodynamic equations, which may be non-trivial in some cases.
For those cases, the soft-constrained approach is a viable, easier-to-use alternative, capable of achieving a level of thermodynamic consistency that is expected to be sufficient for many modeling applications. 
In the soft-constrained approach, using the ALM improves the thermodynamic consistency of predictions and reduces training effort in terms of hyperparameter optimization and epochs required for training.
Its use is therefore always recommended when training in a soft-constrained manner.

Future work should further expand this analysis to more complex mixture properties on experimental data and for numerically challenging target properties.
Additionally, it would be particularly interesting to apply thermodynamics-informed ML models in process design tasks to assess the role consistency of predictions plays in process design and optimization.
This would provide relevant insights to guide further development of ML-based thermodynamic prediction models.
Beyond that, benchmarking thermodynamics-informed GNNs with the recently emerging prior-fitted tabular networks (TabPFN)~\cite{hollmann2025accurate,hicham2026tabular} with respect to thermodynamic consistency may be another interesting research direction.

\section*{CRediT authorship contribution statement}
    \textbf{Jan Pav\v{s}ek:} Conceptualization, Methodology, Software, Formal analysis, Investigation, Writing - original draft, Visualization.
    \textbf{Jan G. Rittig:}
    Methodology, Software, Writing - review \& editing, Funding Acquisition.
    \textbf{Alexander Mitsos:} Conceptualization, Writing - review \& editing, Supervision,  Funding Acquisition.

\section*{Data and Software Availability}

\noindent The data used for training of the presented models are confidential and can be accessed through the NIST Thermodata engine. The code is available in our \emph{GitLab} repository \href{https://gitlab.git.nrw/rwth-avt-svt/public/gmolprop}{\emph{GMoLprop}}.

\section*{Acknowledgements}
\noindent This project was funded by the Deutsche Forschungsgemeinschaft (DFG, German Research Foundation) – 466417970 – within the Priority
Programme ``SPP 2331: Machine Learning in Chemical Engineering''. 

Jan G. Rittig acknowledges funding by the Werner Siemens Foundation within the WSS project of the century “catalaix”.

Model training was performed with computing resources granted by RWTH Aachen University.

The authors thank René Görgen for his software engineering support.

\bibliographystyle{elsarticle-num}  
\bibliography{literature}

\end{document}